\documentclass[11pt]{article}

\usepackage[utf8]{inputenc}
\usepackage[T1]{fontenc}
\usepackage[english]{babel}
\usepackage[caption=false]{subfig}
\usepackage{amsmath}
\usepackage{amssymb}
\usepackage{graphicx}
\usepackage{rotating} 
\usepackage{caption}
\usepackage{float}
\usepackage{geometry}
\usepackage{setspace}
\usepackage{xkeyval}
\usepackage{booktabs}
\usepackage{color}
\usepackage{epstopdf}
\usepackage{authblk}
\usepackage[super,sort&compress,comma]{natbib} 

\usepackage{geometry}
\title{\LARGE{\vspace{-2.0cm}\textbf{Intercalated structures formed by platinum on epitaxial graphene on SiC(0001)}}}

\author[a,$\star$]{Letizia Ferbel}
\author[a]{Stefano Veronesi}
\author[b]{Antonio Rossi}
\author[b]{Stiven Forti}
\author[b]{Camilla Coletti}
\author[a]{Stefan Heun}

\affil[a]{NEST, Istituto Nanoscienze-CNR and Scuola Normale Superiore, Piazza S. Silvestro 12, 56127 Pisa, Italy}
\affil[b]{Center for Nanotechnology Innovation@NEST, Istituto Italiano di Tecnologia, Piazza S. Silvestro 12, 56127 Pisa, Italy}

\affil[$\star$]{\textnormal{letizia.ferbel@sns.it}}

\date{}                     
\begin{document}
\maketitle

\noindent{\textbf{Abstract. }Graphene on SiC intercalated with two-dimensional metal layers, such as Pt, offers a versatile platform for applications in spintronics, catalysis, and beyond. Recent studies have demonstrated that Pt atoms can intercalate at the heterointerface between SiC(0001) and the C-rich $(6\sqrt{3}\times6\sqrt{3})$R30° reconstructed surface (hereafter referred as the buffer layer). However, key aspects such as intercalated phase structure and intercalation mechanisms remain unclear.
In this work, we investigate changes in morphology, chemistry, and electronic structure for both buffer layer and monolayer graphene grown on SiC(0001) following Pt deposition and annealing cycles, which eventually led to Pt intercalation at temperatures above 500~°C.
Atomic-resolution imaging of the buffer layer reveals a single intercalated Pt layer that removes the periodic corrugation of the buffer layer, arising from partial bonding of C-atoms with Si-atoms of the substrate.  In monolayer graphene, the Pt-intercalated regions exhibit a two-level structure: the first level corresponds to a Pt layer intercalated below the buffer layer, while the second level contains a second Pt layer, giving rise to a $(12\times12)$ superstructure relative to graphene. 
Upon intercalation, Pt atoms appear as silicides, indicating a reaction with Si atoms from the substrate.
Additionally, charge neutral $\pi$-bands corresponding to quasi-free-standing monolayer and bilayer graphene emerge. 
Analysis of multiple samples, coupled with a temperature-dependent study of the intercalation rate, demonstrates the pivotal role of buffer layer regions in facilitating the Pt intercalation in monolayer graphene.
These findings provide valuable insight into Pt intercalation, advancing the potential for applications.} \\

\section{Introduction}

Since its discovery, graphene has gathered much attention from researchers worldwide. To date, the growth of graphene from thermal decomposition of silicon carbide (SiC) has emerged as a promising route to scale up the production of high-quality graphene, offering a wafer-scale approach that can meet industrial demands~\cite{Emtsev2009, Mishra2016, Yang2024}. Unlike exfoliation methods, which are limited in scalability, epitaxial graphene formed on SiC provides a more controlled and uniform growth process. One of the key advantages of graphene grown on SiC is its compatibility with existing semiconductor fabrication technologies, making it a more feasible option for integration into devices. 

The epitaxial growth of graphene layers on SiC proceeds via sublimation of Si atoms and the formation of a carbon-rich interfacial layer, also known as buffer layer~\cite{Riedl2010}. The buffer layer is topographically identical to monolayer graphene~\cite{Goler2013}, but about one-third of its carbon atoms are covalently bound to Si atoms from the substrate. This hinders the formation of the typical graphene $\pi$-bands, resulting in the buffer layer being electrically insulating. The buffer layer exhibits a $(6\sqrt{3} \times 6\sqrt{3})$R30° (further referred to as $6\sqrt{3}$) periodicity arising from its interaction and lattice mismatch with the SiC(0001) surface. Despite its promises, the presence of the buffer layer, and thus the coupling with the SiC substrate, leads to a deterioration of the electrical properties of the overlying graphene layers. This significantly limits the practical applications of epitaxial graphene. 

Intercalation of foreign atoms or molecules has proven to be an effective technique to circumvent this problem and to control the electronic properties of graphene~\cite{Qiang2017, Briggs2019, Daukiya2019, Wu2021}. Intercalating foreign species at the buffer/SiC interface allows for decoupling the graphene layer from the substrate, which in turn enhances its mobility and electrical performance~\cite{Robinson2011, Kang2018}. Furthermore, with the intercalation of foreign species, it is possible to modulate the band structure of graphene, induce doping, and form confined 2D phases of the intercalant which find applications in a vast range of fields, from catalysis to sensing, passing through magnetism and superconductivity~\cite{Huempfner2023, Ludbrook2015, Kleeman2013, Kleeman2016, Jeon2013, Girard2023, Filnov2024, Anderson2019, Yang2023, Qiang2017, Briggs2019, Daukiya2019, Wu2021,Jain2020}.

Noble metals (Ag, Au, Pd, Pt) are widely known for their catalytic and photonic properties. Their intercalation has been envisioned as a strategy to protect them from environmental degradation as well as to exploit their properties in a 2D configuration~\cite{Rosenzweig2020, Forti2020, Rybkina2023, Hsu2012}. 
Among noble metals, platinum (Pt) stands out as the best-known catalyst and the element with the largest spin-orbit coupling, owing to its electronic configuration and large mass. These unique characteristics have raised interest across various fields, including catalysis and electronics, particularly spintronics. Previous studies on Pt intercalation in graphene on SiC have demonstrated the formation of quasi-free-standing graphene via Pt intercalation below the buffer layer, as evidenced by diffraction and photoemission measurements~\cite{Xia2014, Idczak2022, Rybkina2023, Weinert2024}. 
However, several key aspects that are highly relevant for the practical application of this material, remain to be clarified. These include the structural details of intercalated phases, precise atomic positioning, and the mechanisms of the intercalation process.

In this work, we use scanning tunneling microscopy (STM) to investigate the changes in morphology and the structure of the intercalated phases obtained after Pt deposition and subsequent annealing, both in buffer layer and monolayer graphene grown on SiC(0001). Additionally, we provide insight into the chemistry and electronic properties of the Pt--intercalated systems via X-Ray Photoelectron Spectroscopy (XPS) and Angle-Resolved Photoemission Spectroscopy (ARPES) measurements. This comprehensive analysis provides a foundation for further advancements toward the potential applications of the system.

\section{Methods}

As substrate for the graphene growth, we employed nominally on-axis n-type 6H--SiC(0001) wafers. The wafers were wet-chemically cleaned and hydrogen etched. The graphene growth was obtained by thermal decomposition of the SiC wafers in a BM-Aixtron reactor at high temperature (1250--1300~°C) under Ar atmosphere~\cite{Emtsev2009, Rossi2018}. With this method, we obtain surfaces consisting of a mixture of buffer-layer (BL) and monolayer graphene (MLG). 

In this manuscript, we employed three samples composed of different buffer layer/graphene ratios: 50/50 (A1), 10/90 (A2), and 90/10 (A3). The graphene quality, uniformity, and composition were first assessed in air by atomic force microscopy and Raman spectroscopy, and later cross characterized by STM analysis in vacuum.  The first two samples (A1 and A2) are discussed within the STM analysis while the third (A3) has been employed for the photoemission spectroscopy measurements. 

Once in the ultra-high vacuum (UHV) chamber of the STM, the as-grown samples were annealed via direct current heating at 600~°C for several hours, followed by 10 min at 800~°C to eliminate adsorbates and to achieve a clean surface. 

Platinum was deposited in the preparation chamber of the UHV-STM at room temperature using a commercial e-beam evaporator. On each investigated sample, Pt was evaporated for 24~min which corresponds to an average Pt thickness of 3.3--3.5~ML. The Pt coverage was calibrated by STM imaging as reported in our previous work~\cite{Ferbel2024}. Pt diffusion and intercalation were achieved by annealing the samples for 10~minutes at temperatures in the range from 500~°C to 800~°C. Temperatures were measured using a K-type thermocouple mounted on the sample holder, directly in contact with the sample, and cross-calibrated with a single wavelength IR-pyrometer with emissivity set to 0.85.

The surface structures were analyzed with a VT-RHK-STM operating in constant current mode, at room-temperature, and under UHV conditions (base pressure $<5\times 10^{-10}$~mbar). STM images were processed with the Gwyddion software package~\cite{Necas2012}. 

The XPS and ARPES measurements were performed in another UHV chamber. Pt deposition and intercalation was carried out in the UHV-STM chamber. Then, the Pt-intercalated samples were transferred in air to the ARPES/XPS chamber where they were outgassed for several minutes at a temperature of $\sim300$~°C.

XPS spectra were acquired using a SPECS XR-50 Mg K$\alpha$ X-ray source and analyzed using the XPST tool implemented in Igor Pro software (WaveMetrics, Inc.). Further details on the fitting procedure are reported in the Supplementary Information (SI). 
ARPES spectra were collected using a He I$\alpha$ radiation (21.2~eV) excitation source (SPECS-$\mu$Sirius) without a monochromator, with a nominal spot size of 100~$\mu$m.

\section{Results \& Discussion}

As reported in earlier findings, platinum atoms, deposited at room temperature, readily form clusters on the epitaxial surface and do not intercalate~\cite{Ferbel2024}.  With annealing cycles above 500~°C, diffusion and intercalation of Pt is achieved. As a consequence, The overall sample morphology drastically changes. Pt clusters adsorbed on the surface agglomerate, and across the sample surface, islands appear that larger but lower than the clusters.

\begin{figure}[htp!]
    \centering
    \includegraphics[width=0.45\linewidth]{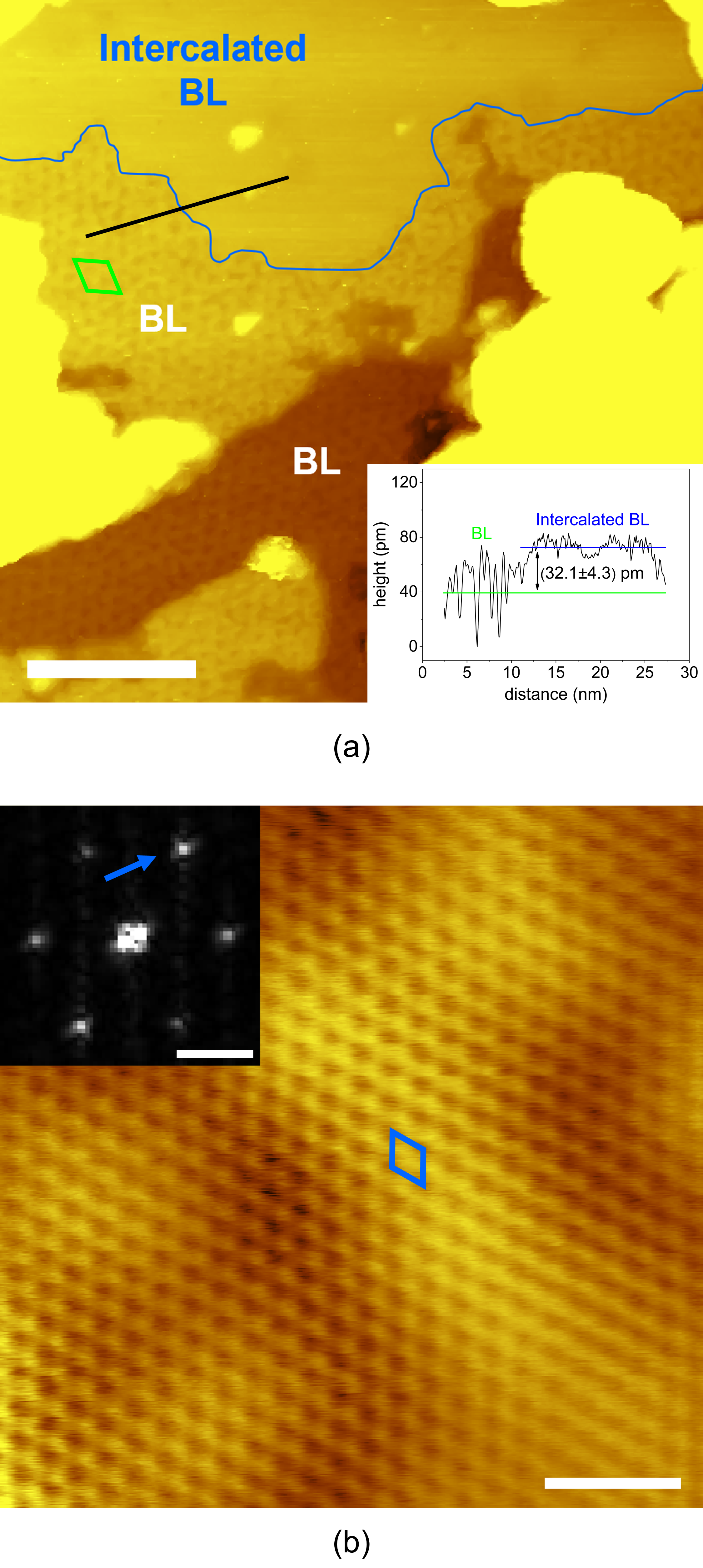}
    \caption{(a) Overview STM scan taken on a buffer layer (BL) region obtained after Pt deposition for 24~min at room temperature and successive annealing up to 600~°C. The contrast-saturated areas correspond to Pt-adsorbed clusters. The green rhombus highlights the $6\sqrt{3}$ unit cell. Inset to (a): Cross-section taken across the line indicated in (a) showing the height difference between intercalated and non-intercalated BL areas. (b) STM topographic image of a Pt-intercalated buffer layer island. The blue rhombus highlights the graphene unit cell. Inset to (b): FFT of the STM image, showing the presence of graphene spots (blue arrow), but no spots related to the $6 \sqrt{3}$ structure. Scale bar: (a) 10~nm, (b) 1~nm, and inset to (b) 4~nm$^{-1}$.}
    \label{Fig1}
\end{figure}

In buffer layer areas, the islands are randomly distributed across the surface, as reported in Fig.~\ref{Fig1}(a), where they are labeled as intercalated BL. They appear as protrusions with strongly reduced corrugation and an apparent height of $(32.1\pm4.3)$~pm with respect to the surrounding non-intercalated buffer layer regions (shown in the inset to Fig.~\ref{Fig1}(a)). The non-intercalated BL regions are clearly identified by the presence of the $6\sqrt{3}$ modulation. The $6\sqrt{3}$ structure arises from the mismatch and partial bonding between the buffer layer and the SiC substrate and is absent in the intercalated regions, as seen in smaller scale STM images (reported in Fig.~\ref{Fig1}(b)), and also confirmed in the FFT of the STM image (inset to Fig.~\ref{Fig1}(b)). This readily indicates that the bonds between the buffer layer and the SiC substrate are no longer present. On the islands, we measure an rms roughness value of $(10.1\pm 1.6)$~pm which is almost half that measured on the buffer layer $(17.9\pm2.1)$~pm. This value is compatible with the rms roughness reported for the buffer layer intercalated by other elements~\cite{Goler2013}. The graphene lattice is resolved over the entire island surface, confirming that Pt is placed below the carbon lattice. These results thus demonstrate that, in the areas of the islands, Pt atoms have intercalated at the buffer layer/SiC interface and transformed the buffer layer into quasi-free-standing (QFS) monolayer graphene (i.e., the islands correspond to \mbox{MLG/Pt/SiC}). 

\begin{figure}[ht]
    \centering    \includegraphics[width=0.7\linewidth]{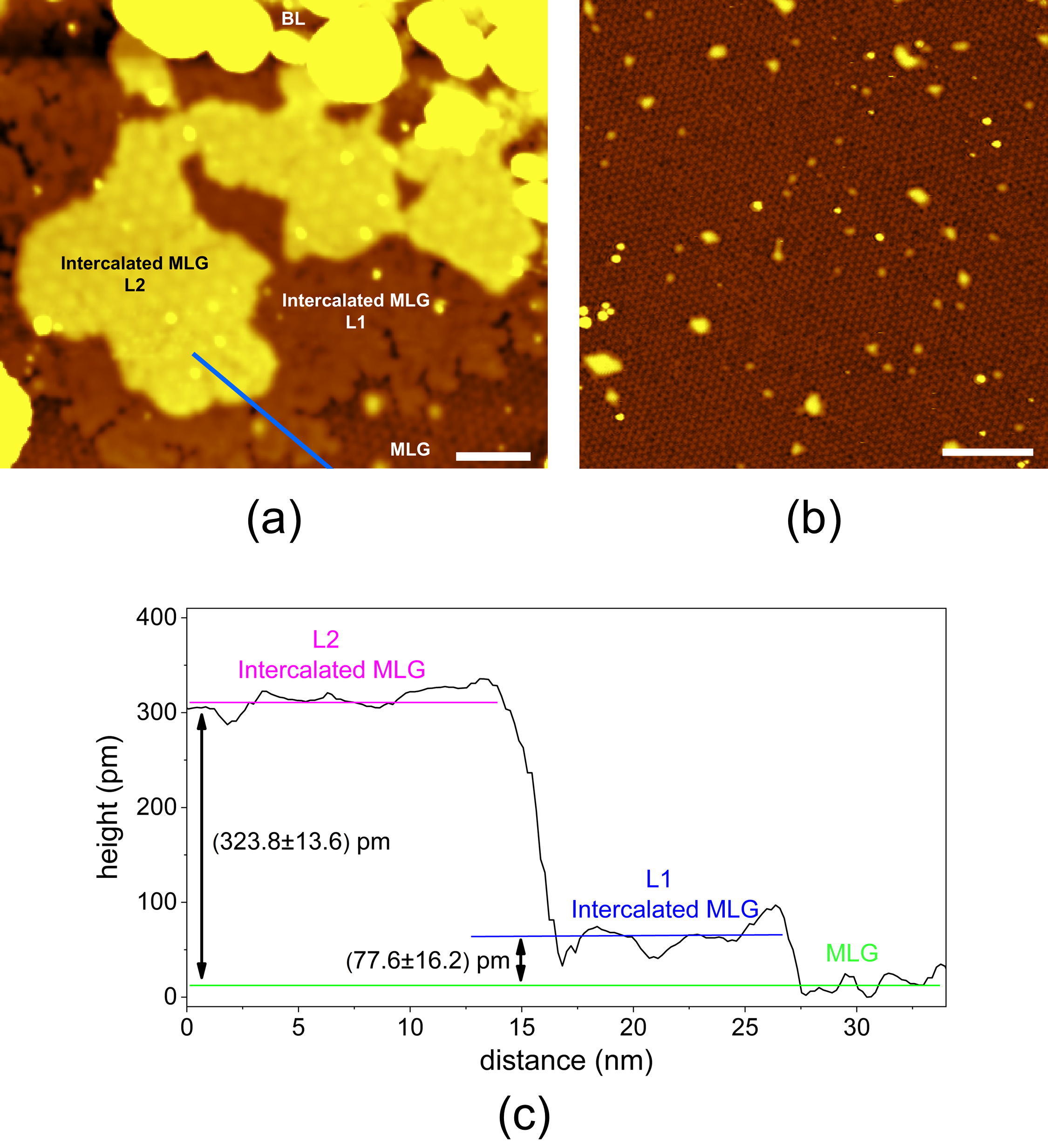}
    \caption{Overview STM scans taken on the monolayer graphene (MLG) surface obtained after Pt deposition for 24~min and annealing up to 600~°C. Scan in (a) taken in an area close to the buffer layer while (b) taken in the middle of a terrace. (c) Cross-section, taken across the line indicated in (a), showing the height difference between intercalated levels L1 and L2 and non-intercalated MLG areas. Scale bar: (a) 10~nm and (b) 20~nm.}
    \label{Fig2}
\end{figure}

In monolayer graphene regions, islands are observed as well. These islands are mainly scattered at the boundaries with the buffer layer (as reported in Fig.~\ref{Fig2}(a)), which suggests that these facilitate the Pt intercalation process. 
However, islands are also found randomly distributed within the terraces (as shown in Fig.~\ref{Fig2}(b)). In this case, Pt intercalation is likely initiated by defects in the honeycomb lattice, pre-existing or formed during the process, and the intercalation below the buffer layer proceeds via a two-step diffusion process i.e., penetration of the graphene membrane and further diffusion below the buffer layer. 
In monolayer graphene areas, two--levels of islands are observed. We label the lower level as "L1", and the higher level as "L2", as reported in Fig.~\ref{Fig2}(a). These levels have an apparent height difference with respect to the surrounding non-intercalated monolayer graphene areas of $(77.6\pm16.2)$~pm (L1) and $(323.8\pm13.6)$~pm (L2), as reported in Fig~\ref{Fig2}(c). 

\begin{figure}[htp!]
    \centering    \includegraphics[width=0.45\linewidth]{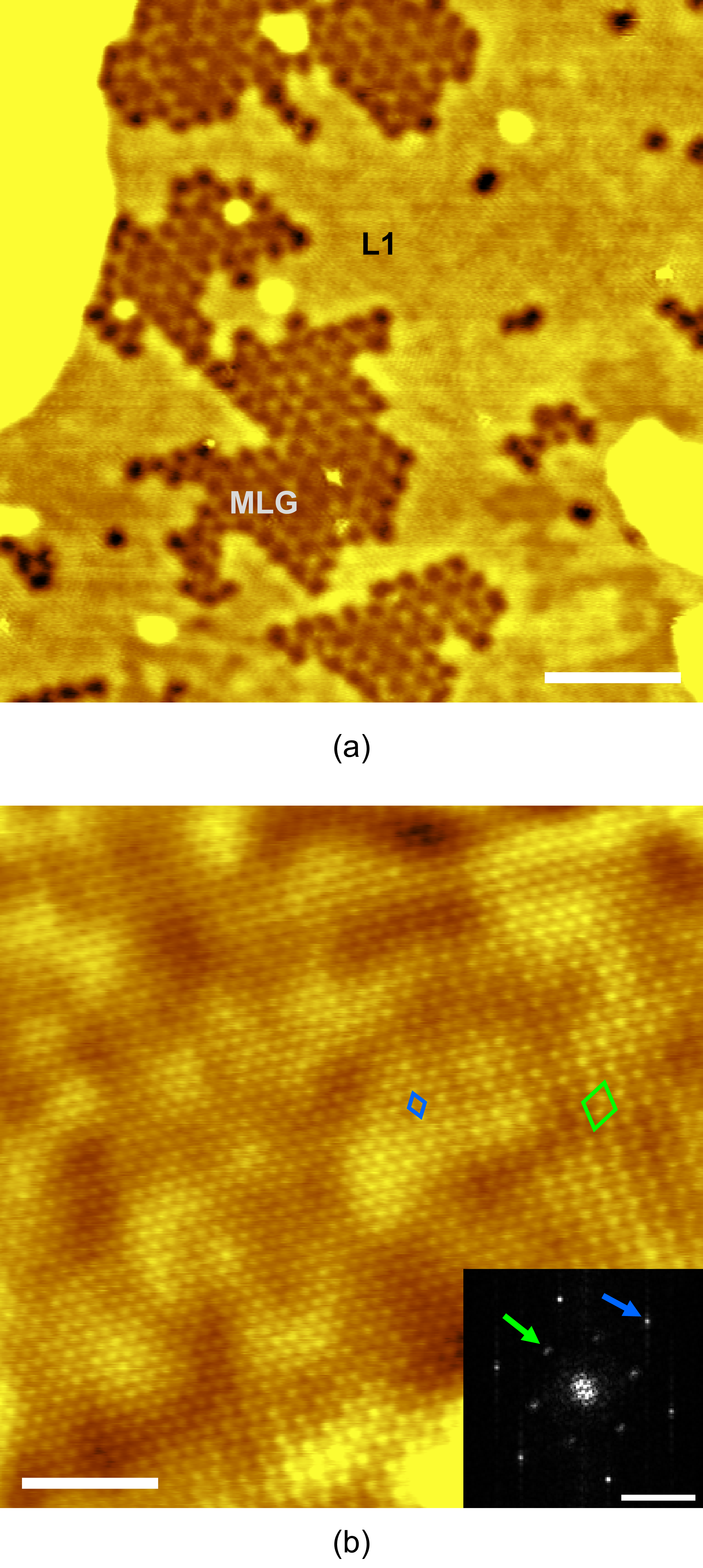}
    \caption{STM topographic images of Level 1. (a) Overview STM scan imaging Level 1 (L1) of Pt-intercalated graphene together with a clean monolayer graphene area (MLG). The contrast-saturated white areas correspond to Level 2 (L2). (b) Close-up view of Level 1 showing the honeycomb lattice of graphene together with the $(\sqrt{3}\times\sqrt{3})$R30° structure. The two unit cells are highlighted in blue and green. Inset to (b) FFT of the STM scan in (b) showing maxima induced by the graphene lattice (blue arrow) and by the $(\sqrt{3}\times\sqrt{3})$R30° modulation (green arrow). Scale bar: (a) 10~nm, (b) 2~nm, and inset to (b) 4~nm$^{-1}$.}
    \label{Fig3}
\end{figure}

Figure~\ref{Fig3} shows STM scans taken from the lower level (L1) of Pt intercalated islands in the monolayer graphene region (as labeled in Fig.~\ref{Fig2}(a)). The first striking feature is the absence of the $6\sqrt{3}$ modulation in L1, while the $6\sqrt{3}$ is still clearly observable in non-intercalated areas. This is again indicative of the absence of bonding between the buffer layer and the SiC substrate and suggests intercalation of Pt under the buffer layer. The surface of L1 is quite flat with an rms roughness of $(8.6\pm2.4)$~pm, which is almost half of that of pristine monolayer graphene ($(13.7\pm 1.9)$~pm) and comparable to the value measured on the Pt--intercalated buffer layer islands. The graphene lattice is resolved over the entire L1 surface, confirming that Pt is placed below the graphene lattice. In the L1 areas, a long-range ordered $(\sqrt{3}\times\sqrt{3})$R30°pattern with respect to the graphene lattice (further referred to as $\sqrt{3}$) is observed in STM. This is confirmed in the FFT of the STM image (inset to Fig.~\ref{Fig3}(b)). The unit cells of graphene and the $\sqrt{3}$ are highlighted in blue and green, respectively, in Fig.~\ref{Fig3}(b). 

The $\sqrt{3}$ pattern has often been reported and interpreted as formed by metal atoms sandwiched between graphene layers~\cite{Fiori2017, Watcharinyanon2011}. Since we observe the lifting of the $6\sqrt{3}$ reconstruction and thus Pt intercalation under the buffer layer, this interpretation would imply the presence of two platinum layers: one at the buffer/SiC interface and one between the buffer and the graphene layer. However, the measured apparent height difference between L1 and the surrounding monolayer graphene is only compatible with a single platinum layer. Thus, this interpretation for the $\sqrt{3}$ modulation does not fit the data. The emergence of a $\sqrt{3}$ pattern has also been observed after intercalation at the buffer layer-SiC interface~\cite{Kotsakidis2020,Rosenzweig2019,Link2019, Toyama2022}. Recently, the $\sqrt{3}$ pattern was attributed to a quantum interference effect~\cite{Schadlich2023, Qu2022}, which is likely what happens also here. Since Pt intercalation in monolayer graphene proceeds either at phase boundaries or defects, either preexisting or generated by the Pt intercalation itself, the appearance of the $\sqrt{3}$ is most likely mediated by the presence of such defects in the graphene lattice.
This interpretation of quantum interference is also consistent with the fact that the $\sqrt{3}$ pattern was only seen in the monolayer areas and not in the BL. Contrary to MLG, the Pt intercalation process in the BL does not increase the number of defects but rather removes them. Additionally, this interpretation only requires the presence of a single Pt layer at the buffer/SiC interface. In summary, the absence of the $6\sqrt{3}$ modulation, the reduced rms roughness, and the relatively small height difference between L1 and the surrounding monolayer graphene thus suggest that in L1 there is a single layer of Pt atoms which is intercalated between the buffer layer and the SiC substrate, thus QFS-BLG is formed (i.e., L1 of the islands corresponds to MLG/MLG/Pt/SiC). 

\begin{figure}[htpb!]
    \centering
    \includegraphics[width=\linewidth]{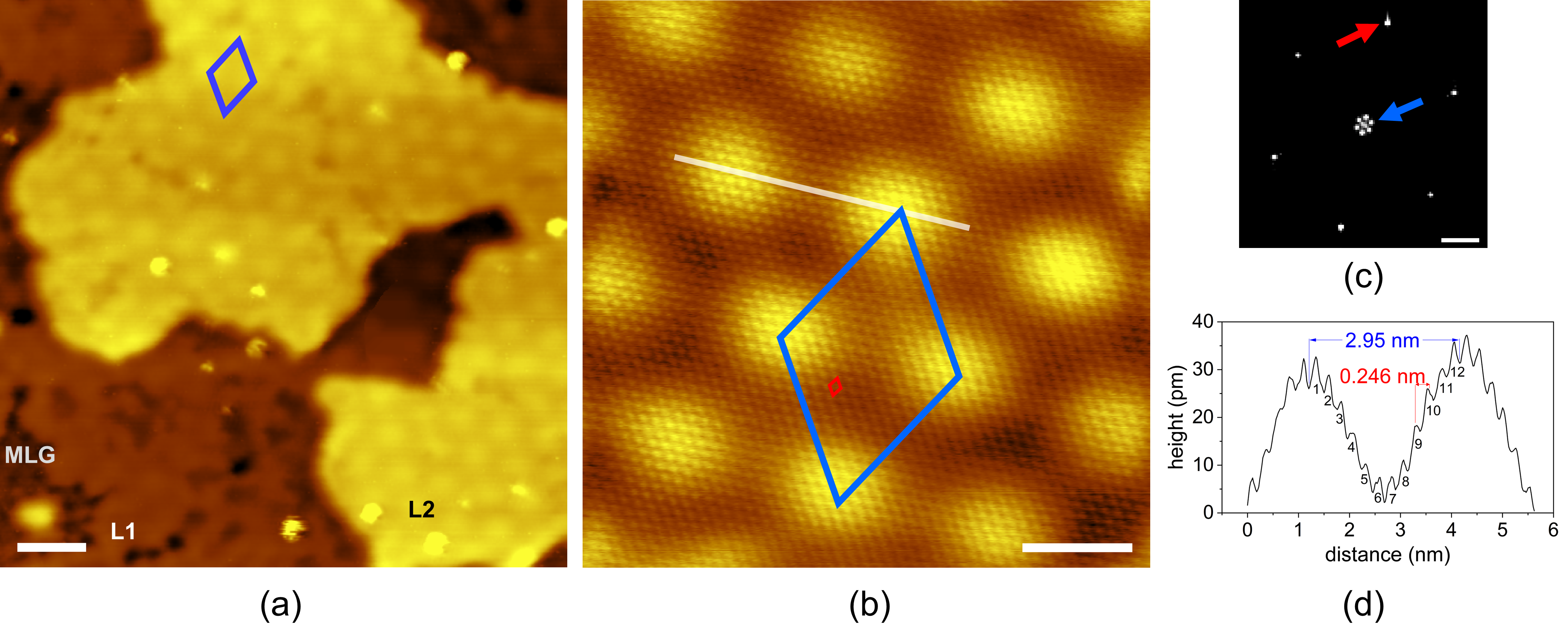}
    \caption{STM topographic images of Level 2 (L2). (a) Overview STM scan showing level L2 of Pt-intercalated graphene together with level L1 and a clean monolayer graphene (MLG) area. The blue rhombus highlights a unit cell of the Pt-induced moiré in L2. (b) Close-up view of level L2 showing the honeycomb lattice of graphene together with the $(12\times12)$ moiré pattern. The two unit cells are highlighted in blue and red, respectively. (c) FFT of the STM scan in (b) showing maxima induced by the graphene lattice (red arrow) and by the $(12\times12)$ moiré pattern (blue arrow). (d) Cross section taken along a unit cell vector of the moiré pattern, as indicated by the line in (b), highlighting the presence of 12-peaks (i.e., 12-graphene unit cells) between two maxima of the moiré pattern. Scale bar: (a) 5~nm, (b) 2~nm,  and (c) 2~nm$^{-1}$.}
    \label{Fig4}
\end{figure}

Figure~\ref{Fig4} shows STM scans taken from the upper level (L2) of Pt intercalated islands in the monolayer graphene region (as labeled in Figs.~\ref{Fig2}(a)). The $6\sqrt{3}$ structure is no longer observed. Instead, we see a regular hexagonal arrangement of bright protrusions with a periodicity of $(2.92\pm0.17)$~nm, highlighted in Fig.~\ref{Fig4}(a-b). This pattern can be understood as a moiré pattern arising from the interaction between the graphene lattice and the Pt layer. As shown in Fig.~\ref{Fig4}(b), the graphene lattice and this moiré superstructure can be imaged together. The unit cells of the two structures are highlighted in red and blue, respectively. We observe that the two unit cells have a 0° twist angle, as also confirmed by the FFT of the STM image reported in Fig.~\ref{Fig4}(c). The line scan shown in Fig.~\ref{Fig4}(d), taken between two moiré protrusions, shows that the moiré periodicity corresponds to 12 graphene maxima, i.e., 12 graphene unit cells. This implies that the $2.92\pm0.17$~nm moiré structure is compatible with a $(12\times12)$ superstructure with respect to graphene ($0.246\text{~nm}\times12 = 2.952\text{~nm}$ period). 

Again, the absence of the $6\sqrt{3}$ indicates the presence of Pt atoms intercalated at the buffer layer/SiC interface. However, since on L2 we measure an apparent height difference $(244.8\pm9.7)$~pm with respect to L1, it is reasonable to assume the presence of a second Pt layer which gives rise to this height difference. Since we observe the graphene lattice extending all over the topmost graphene surface, for this second platinum layer there are only two possible locations: either below the buffer layer (thus forming a Pt bilayer at the buffer/SiC interface) or between graphene and buffer layer. We observed a two-level structure only in monolayer graphene regions. In buffer layer regions, we only observed single level islands that show L1-like arrangement. Additionally, Pt intercalation below the buffer layer is expected to generate, if any, a moiré pattern that is either commensurate or has an epitaxial relationship with the underlying SiC substrate rather than with the graphene lattice~\cite{Ciesler2023}. Since we observe a moiré pattern that is aligned and commensurate to the graphene lattice, there must be a Pt layer, intercalated between the graphene and the buffer layer, arranged in a closely-packed-hexagonal structure with the unit cell aligned to that of graphene. The Pt-Pt distance in a hexagonal (111) plane of bulk Pt is 0.2775~nm~\cite{Pearson1958}. This suggests that the $(12\times12)_{Gr}$ moiré pattern corresponds to an $(11 \times 11)_{Pt}$ structure in which Pt atoms are arranged in a hexagonal pattern and have a Pt-Pt distance of $(0.266 \pm 0.005)$~nm, i.e., the Pt atoms in this layer are $\approx 4\%$ laterally compressed compared to Pt atoms in the (111) surface (0.2775~nm~\cite{Pearson1958}). To further notice that the height value of $(244.8\pm9.7)$~pm measured for this interlayer is only slightly larger than the interplanar spacing of Pt(111) which is 226.5~pm~\cite{Pearson1958}. The in-plane compression of the Pt layer would likely lead to a strain release in the perpendicular direction, thus leading to a larger interplanar spacing.

Thus the L2 islands are due to the presence of an additional Pt layer with respect to L1. L2 comprises two Pt layers, with one of them intercalated at the buffer layer/SiC interface (i.e., L1) giving rise to QFS-bilayer graphene (BLG) and a second layer arranged in a hexagonal lattice sandwiched between the former buffer layer and the monolayer graphene, which gives rise to the $(12\times12)_{Gr}$ moiré pattern (i.e., L2 corresponds to \mbox{MLG/Pt/MLG/Pt/SiC}). 

\begin{figure}[htpb!]
    \centering
    \includegraphics[width=0.7\linewidth]{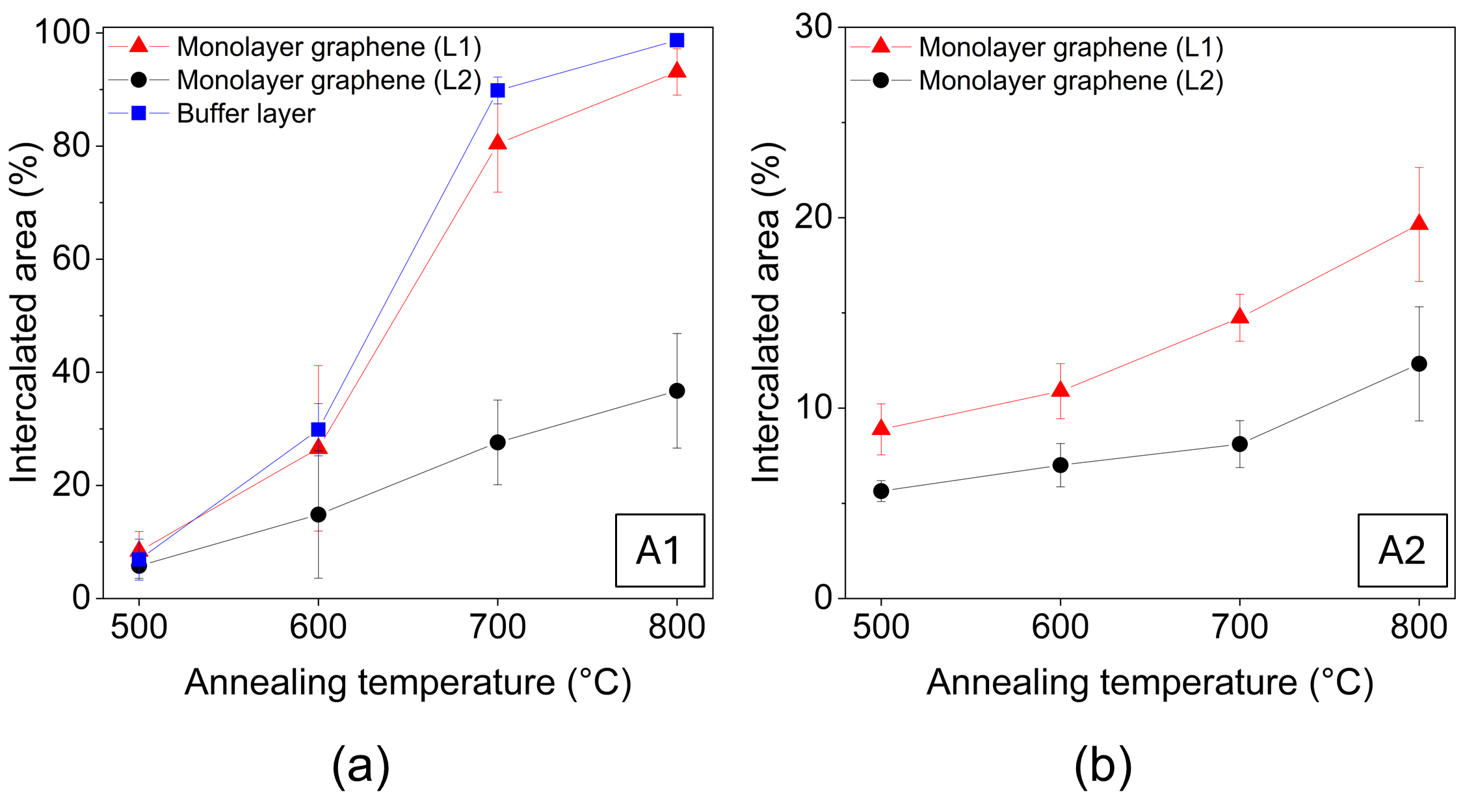}
    \caption{Statistical analysis of large-scale STM images reporting the surface fraction occupied by the intercalated Pt islands as a function of the annealing temperature. The lines connecting the data points are shown as a guide to the eye. Analysis performed (a) on sample A1 and (b) on sample A2.}
    \label{Fig5}
\end{figure}

Further insight into the intercalation process can be gained by analyzing the extension of the intercalated regions as a function of the annealing temperature. Note that the morphology of the intercalated islands did not change within the temperature range studied (500~°C--800~°C).
The graphs in Fig.~\ref{Fig5} report the evolution of the intercalated area fraction of two samples containing a different amount of buffer layer, onto which we deposited the same amount of Pt and which were subjected to annealing cycles from 500~°C to 800~°C.

For sample A1 (50/50 BL/MLG ratio), complete intercalation of Pt atoms at the buffer/SiC interface was achieved by annealing cycles up to 800~°C (100\% intercalation), as shown in Fig.~\ref{Fig5}(a). The rather abrupt increase in the intercalated area fraction for the annealing cycles from 600~°C to 700~°C suggests the presence of an activation energy barrier for the Pt intercalation process. Interestingly,  buffer layer and monolayer graphene show the same intercalation growth trend. This suggests that the two intercalation processes require similar activation energies. On the other hand, for sample A2 (10/90 BL/MLG ratio), Pt atoms intercalated only a small fraction of the monolayer graphene, as shown in Fig.~\ref{Fig5}(b). After annealing at 800~°C, only $\sim$20\% of the monolayer graphene became QFS-BLG.

In buffer layer regions, with the annealing cycles, the Pt clusters adsorbed on the surface only slightly agglomerate and, thanks to the availability of Pt atoms all across the BL surface, Pt intercalation is initiated at random locations of the BL surface. The intercalated islands enlarge with increasing annealing temperature until at 800~°C Pt atoms intercalate the entire surface, at which point all the buffer layer has turned into QFS-MLG. As the intercalated islands enlarge in size, the Pt clusters that are still adsorbed on the surface reduce in size and move to non-intercalated buffer layer areas.

In monolayer graphene regions, with the annealing cycles, the Pt clusters more strongly agglomerate and mostly accumulate at the step edges and at the boundaries with the buffer layer. This limits the availability of Pt atoms to intercalate at random locations on the surface.
Indeed, Pt intercalated islands in MLG far from step edges and boundaries with the buffer layer are rather small and dispersed (see Fig.~\ref{Fig2}). With increasing annealing temperature, these intercalated islands increased in number but only slightly in size. Indeed, they did not exceed a lateral size of 20~nm. 
Conversely, close to the boundaries with the buffer layer, intercalated islands in MLG readily grow, and with increasing annealing temperature they enlarge, incorporating the small islands within the terrace, until full intercalation of the monolayer graphene is reached. 
This behavior explains why full intercalation of monolayer graphene was only obtained in samples comprising large amounts of buffer layer. 

\begin{figure}[htpb!]
    \centering
    \includegraphics[width=\linewidth]{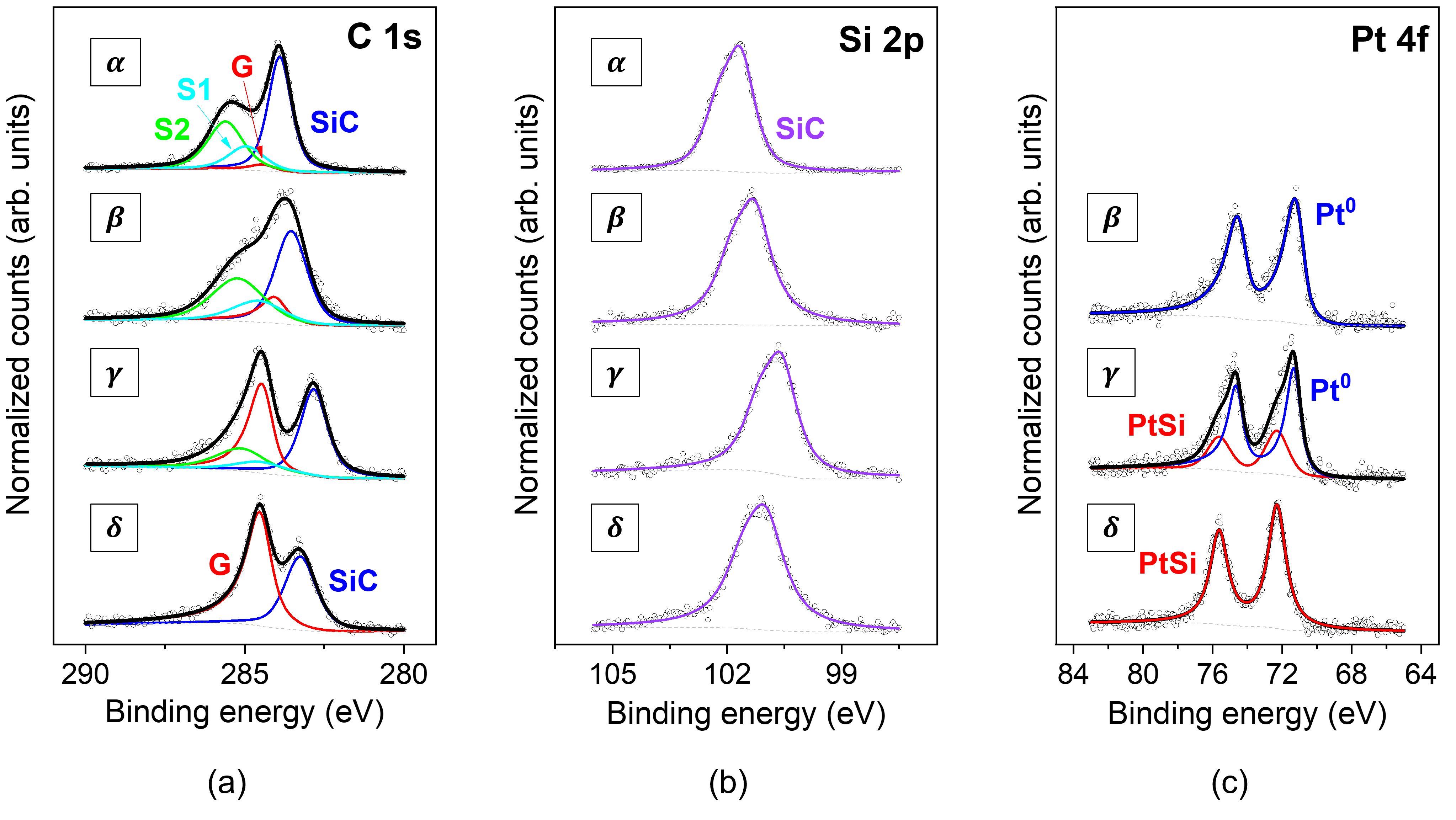}
    \caption{(a) C~1s, (b) Si~2p, and (c) Pt~4f core level spectra from the initial graphene sample ($\alpha$), after Pt deposition at RT ($\beta$), and subsequent annealing at 600~°C ($\gamma$) and 800~°C ($\delta$). Individual components are indicated (and labeled) in each spectrum.}
    \label{Fig6}
\end{figure}

Intercalation of the buffer layer via Pt intercalation is further confirmed by XPS measurements shown in Fig.~\ref{Fig6} (a summary of the fitting parameters is given in Tables~S1--S3 in the SI), which shows the evolution of the C~1s, Si~2p, and Pt~4f core level spectra for sample A3 (90/10 BL/MLG ratio) obtained after Pt deposition at RT, followed by annealing cycles at 600~°C and 800~°C.

The C~1s spectrum of the pristine sample consists of four components: S1 (284.95~eV) and S2 (285.6~eV) correspond, respectively, to sp$^3$ and sp$^2$ C atoms in the buffer layer. The SiC component (283.9~eV) corresponds to C atoms of the SiC substrate, and the small low-intensity G component (284.5~eV) originates from C atoms in graphene~\cite{Emtsev2008}. The Si~2p peak presents a single spin-orbit doublet with the $2p_{3/2}$ component located at 101.64~eV. 

After Pt deposition, the Pt core-level spectrum shows a single asymmetric 4f doublet with the 4f$_{7/2}$ spin-orbit component located at 71.4~eV, corresponding to metallic Pt~\cite{Wertheim1988, Moulder1992}. The C~1s and Si~2p core levels rigidly shift by $\sim0.4$~eV to lower binding energies, suggesting a band bending deriving from Pt adatoms on the surface. Previous works reported only little to no shift in C~1s or Si~2p envelopes~\cite{Xia2014, Idczak2022}. However, this difference compared to our results is likely due to the higher buffer layer content or the significantly greater Pt coverage used in our study.

All core levels change significantly after annealing the Pt-deposited sample at 600~°C. An additional symmetric 4f doublet (4f$_{7/2}$ at 72.3~eV) appears in the Pt~4f core-level spectrum located 0.9~eV to higher binding energies than the metallic Pt peak. This position is characteristic of platinum silicides~\cite{Moulder1992} and readily suggests that a reaction between Pt atoms and Si atoms from the SiC substrate has occurred indicating intercalation of Pt atoms below the buffer layer. In the C~1s core level spectrum, the graphene sp$^2$ component (284.52~eV) is strongly increased, indicating that in some regions the buffer layer has been decoupled from the substrate and turned into quasi-free-standing graphene. At the same time, the intensity of the buffer layer decreases but does not disappear. The SiC substrate component now appears at 282.83~eV, corresponding to a $\sim1.1$~eV ($\sim0.7$~eV) shift to lower binding energies compared to the position of the SiC component in the pristine sample (in the as-deposited sample). A consistent shift of $\sim1.1$~eV ($\sim0.7$~eV) to lower binding energies is also observed in the Si~2p core level spectrum, which now shows the 2p$_{3/2}$ spin-orbit component at 100.57~eV. This shift in the SiC substrate component can be associated with a change in the dipole layer at the buffer layer/SiC interface caused by the presence of Pt atoms at the interface~\cite{Watcharinyanon2012}. 

With further annealing of the sample to 800~°C, full decoupling of the buffer layer is achieved. The Pt~4f core-level spectrum shows the sole presence of the higher binding energy 4f doublet (4f$_{7/2}$ at 72.3~eV) associated with platinum silicide. The C~1s core level spectrum is characterized by two components corresponding to graphene (284.59~eV) and SiC (283.25~eV), indicating full decoupling of the buffer layer from the substrate. The SiC component shows a shift of $\sim0.65$~eV to lower binding energies compared to the pristine sample, also observed in the Si~2p core level spectrum (2p$_{3/2}$ at 100.98~eV). This shift is compatible with a change of the surface chemistry at the buffer/SiC interface induced by Pt atoms bonded to the topmost silicon layer of the SiC substrate. 

It is interesting to notice the shift of $\sim0.4$~eV observed in the SiC components between the 600~°C and 800~°C annealing cycles. Similarly, Weinert et al.~\cite{Weinert2024} reported two shifted SiC components coexisting in Pt intercalated BL samples. In our study, we might attribute the presence of different SiC shifted components to the bonding state and location of the Pt atoms. After annealing the Pt--deposited sample at 600~°C, the Pt~4f spectrum shows the presence of metallic Pt adatoms together with intercalated Pt atoms which have formed silicides at the interface. At this intercalation step, the buffer layer surface thus comprises regions with Pt adatoms and regions of QFS-MLG. Such arrangement of Pt atoms causes band bending at the surface or/and at the interface. 
On the other hand, after annealing the sample at 800~°C, in the Pt~4f spectrum, only the silicide component can be resolved. All buffer layer area turned into QFS-MLG. Thus, Pt atoms are expected to cause only band bending at the interface region. 

\begin{figure}[hb!]
    \centering
    \includegraphics[width=0.7\linewidth]{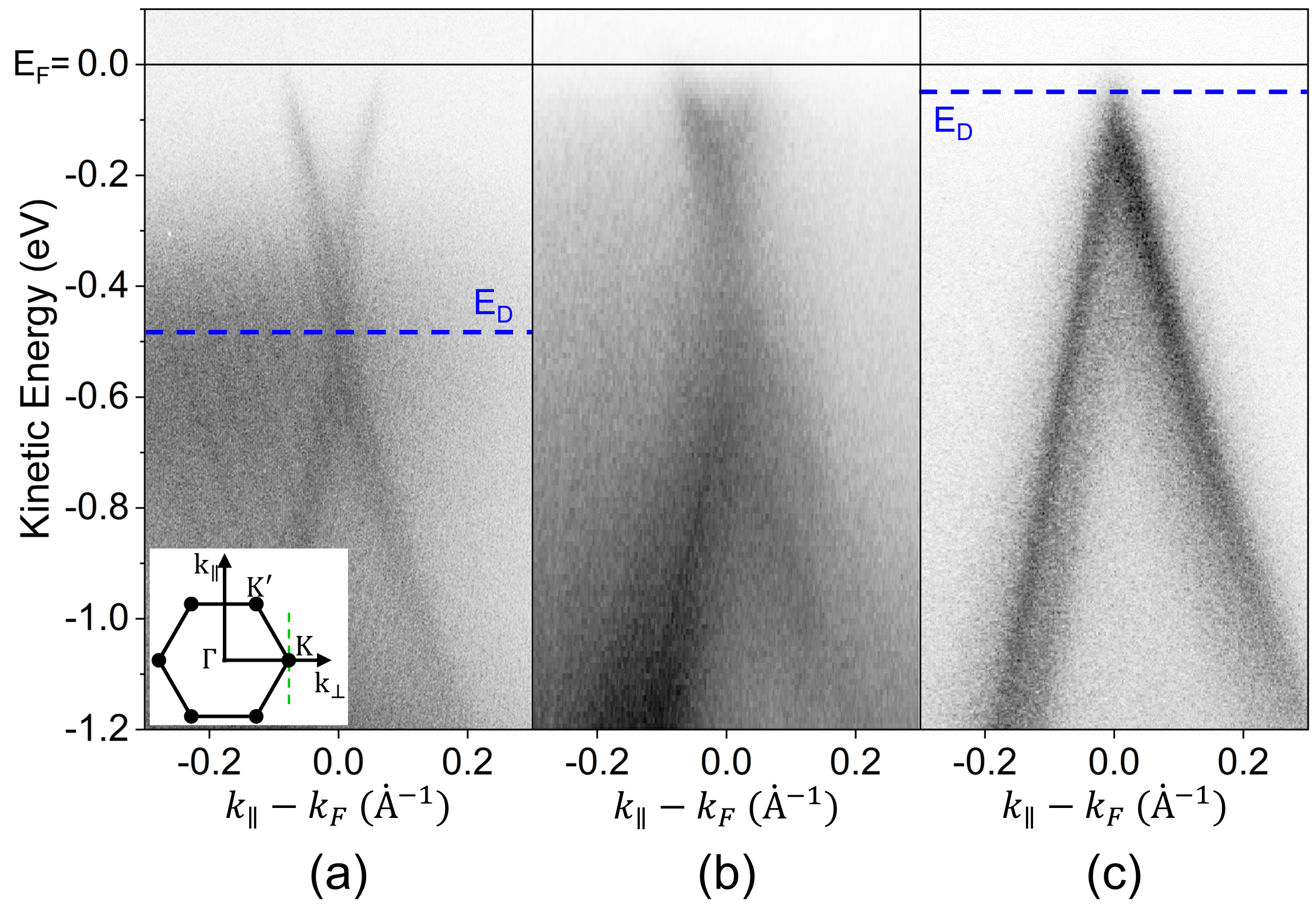}
    \caption{ARPES spectra around the graphene K-point of the Brillouin zone (see the inset in (a)) obtained from (a) the pristine sample A3, and after Pt deposition and annealing at (b) 600~°C and (c) 800~°C. The dashed blue line indicates the position of the Dirac point.}
    \label{Fig7}
\end{figure}

The final evidence of successful, step-wise, Pt intercalation with increasing annealing cycles is provided by ARPES measurements performed on sample A3 around the graphene K-point of the Brillouin zone (reported in Fig.~\ref{Fig7}). 

The band dispersion of the pristine sample (shown in Fig.~\ref{Fig7}(a)) shows the typical graphene $\pi$-band dispersion, although weak. This arises from the small fraction of monolayer graphene present in the sample. The Dirac point (E$_D$) is located at around $-0.48$~eV below the Fermi level (E$_F$), as typically observed for epitaxial graphene on SiC~\cite{Forti2017}. Besides, the data shows two non-dispersive states located at around 0.4~eV and at 1.2~eV, arising from the BL (highlighted in Fig.~S3 in the SI). 

After Pt deposition and annealing to 600~°C, the ARPES spectrum in Fig.~\ref{Fig7}(b) shows a rather high background caused by inhomogeneous scattering. Nevertheless, multiple diffuse $\pi$-bands can be observed, arising from partial passivation of the buffer layer by Pt intercalation.

Further annealing the sample at 800~°C leads to a significant decrease in the background intensity (as shown in Fig.~\ref{Fig7}(c)) and the complete disappearance of the non-dispersive buffer layer states (as shown in Fig.~S3 in the SI), indicating complete intercalation of the sample. The $\pi$-bands become sharper and a weak, yet visible, bilayer band structure is detected as well. The Dirac point is sharply shifted towards the Fermi level, indicating the formation of quasi-free-standing graphene~\cite{Xia2014,Rybkina2023,Riedl2009}.

\section{Conclusions}

We have reported a detailed analysis of the changes in buffer layer and monolayer graphene grown on SiC(0001) induced by Pt deposition and subsequent intercalation. 
With STM imaging, we provide the first real-space characterization of the Pt intercalation morphology. In buffer layer regions, a single layer of Pt atoms intercalates below the buffer layer, disrupting the $6\sqrt{3}$ superstructure caused by the interaction between the buffer layer and SiC substrate, but without forming any additional periodicity. In monolayer graphene areas, Pt intercalated regions show up as a two-level structure with one Pt layer intercalated below the buffer layer and a second Pt layer sandwiched between the former buffer layer and monolayer graphene. This second Pt layer is arranged in an ordered structure which gives rise to a moiré pattern corresponding to a $(12\times12)$ superperiodicity with respect to graphene.
Furthermore, STM analysis shows that for Pt intercalation below monolayer graphene, the presence of buffer layer/monolayer graphene boundaries plays a pivotal role. Indeed, full decoupling of monolayer graphene could not be obtained in the absence of sizeable buffer layer regions in the epitaxial samples. 
Through XPS we have further shown that the decoupling of the buffer layer from the substrate is linked to the formation of Pt silicides at the topmost SiC surface, consistent with previous reports~\cite{Xia2014, Idczak2022, Rybkina2023, Weinert2024}.
The disappearance of buffer layer non-dispersive states and progressive formation of almost charge-neutral single and bilayer-like $\pi$-band dispersions were observed in ARPES. 
In conclusion, using STM, XPS, and ARPES we demonstrate that Pt atoms intercalate below the buffer layer and effectively decouple the latter from the SiC substrate.

\section*{Acknowledgements}
We acknowledge financial support from the PNRR MUR Project PE0000023-NQSTI founded by the European Union-NextGenerationEU.

\bibliography{bibliography} 
\bibliographystyle{ieeetr} 
\end{document}